\documentclass{epl}

\usepackage{amsmath,amsfonts,amssymb}
\usepackage{graphicx}
\usepackage{epsfig}

\title{An algorithm for counting circuits: application to real-world
and random graphs.}
\shorttitle{An algorithm for counting circuits}
\author{E.~Marinari\inst{1} \and R.~Monasson\inst{2} \and G.~Semerjian\inst{3}}
\institute{
\inst{1} Dipartimento di Fisica and INFN, Universit\`a di Roma La Sapienza, 
P. A. Moro 2, 00185 Roma, Italy.\\
\inst{2} CNRS-Laboratoire de Physique Th\'{e}orique de 
l'ENS, 24 rue Lhomond, 75005 Paris, France,\\
\inst{3}
Dipartimento di Fisica and SMC-INFM, Universit\`a di Roma La Sapienza, 
P. A. Moro 2, 00185 Roma, Italy.
}
\begin{document}

\maketitle

\begin{abstract} 
We introduce an algorithm which estimates the number of circuits in a
graph as a function of their length. This approach provides analytical results 
for the typical entropy of circuits in  sparse random graphs. When applied to
real-world networks, it allows to estimate exponentially large numbers of 
circuits in polynomial time. We illustrate the method by studying a graph of
the Internet structure.
\end{abstract}

{\em Introduction.}  An increasing amount of data has been collected
on the topology of real-world networks appearing in many different
contexts, the Internet being only one of many
examples~\cite{review_datas}. A natural line of research in this field
consists in identifying characteristic features of the networks, to
compare them with theoretical models and potentially
disprove the latter.  The simplest of these properties is the
distribution of vertex degrees, which has been repeatedly argued to
exhibit power-law tails. Quite generally, it is computationally easy
to measure the `local' (involving a vertex and a finite number of
neighbors) properties of a given network; for instance loops with less
than $5$ edges in the Internet graph have been studied
in~\cite{sl_exp}. However, it might well be that the most distinctive
features of real-world networks are `global' ({\em i.e.} that they depend on
an extensive portion of the graph): their measure becomes then a very
challenging numerical problem.

Among these global properties, we shall consider in the present work
the number of long circuits in a graph, {\em i.e.} of circuits
visiting a finite fraction of the vertices~\cite{ben}. 
These circuits are roughly
exponentially numerous in the size of the graph: because of that the
use of exact algorithms~\cite{Johnson}, whose complexity is linear in
the number of cycles to be enumerated, is possible for small sizes
only.  Another trace of this difficulty lies in the NP-completeness of
the decision problem of knowing if a graph is Hamiltonian ({\em i.e.} if it
contains a loop visiting all vertices)~\cite{complexity}. To overcome
this difficulty it is reasonable to look for approximate algorithms
with running times scaling polynomially with the network size.  A
formal result in this direction is the existence of a probabilistic
algorithm for the approximate counting of Hamiltonian cycles in graphs
with large minimal connectivity~\cite{proba_Hamilton}. Very recently a
sampling method based on a Monte Carlo Markov Chain has been
proposed~\cite{MC_enumeration}.

In this letter we introduce a counting algorithm, relying on a
statistical mechanics approach expanding on the results of~\cite{MaMo}
(see also~\cite{polymers}), the details of which shall be exposed in a
longer publication~\cite{long}.  The algorithm is first applied to a
real-world network, then to random graphs. In the latter context the
number of circuits of a given length is a random variable, whose 
properties have been thoroughly
studied by Garmo~\cite{Garmo} in the regular case (all vertices have the
same degree, see also~\cite{Wormald_review} for a review). 
For the Erd\"os-R\'enyi ensemble (in which vertex degrees
are Poisson distributed), most results have been obtained in the
vicinity of the percolation transition, or on the contrary for very
large average degrees~\cite{Bollo_book}. The
calculation of the expected number of circuits in ensembles with
arbitrary degree distributions has been recently performed by Bianconi
and Marsili~\cite{BiMa}.  However these expectations turn out to be
dominated by atypical (exponentially rare) graphs with exponentially more
circuits than the typical ones. 

\begin{figure}
\begin{center}
\includegraphics[width=3.8cm]{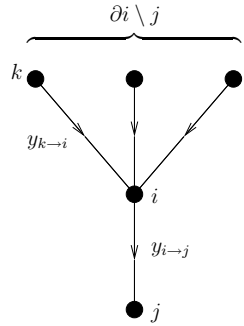}
\end{center}
\caption{A vertex $i$, and its neighbors $j$ and $ k \in \partial
  i\setminus j$. Each oriented edge carries a message $y$ 
involved in Eq.~(\ref{eq_msgs}).}
\label{fig_msgs}
\end{figure}

{\em Definitions and algorithm.}  
Let $G=(V,E)$ be a graph, where $V$ and $E$ are the sets of vertices
and edges respectively. The size of $G$ is the number of vertices,
$N=|V|$. A circuit ${\cal C}=(V_{\cal C},E_{\cal C})$ is a closed path
on the graph visiting each vertex at most once. The length $L$ of the
circuit is the number of visited vertices or edges, $L=|V_{\cal C}| =
|E_{\cal C}|$. A circuit visiting all vertices ($L=N$) is called
Hamiltonian. Our scope is to count the number of distinct circuits of
a given graph $G$, ${\cal N}_L(G)$, as a function of their length
$L$; more precisely, we define below a procedure estimating the
entropy $\sigma(\ell) =(\ln {\cal N}_L )/N$ of circuits of length
$N\,\ell$.  The reduced length $\ell$ is an intensive parameter in
$[0,1]$.  For $i\in V$, we call $\partial i$ the set of neighbors of
the vertex $i$, and use the symbol $\setminus$ to subtract an element
of a set: if $j$ is a neighbor of $i$, $\partial i \setminus j$ will
be the set of all neighbors of $i$ distinct from $j$. We denote by
$i\to j, j\to i$ the two oriented edges that can be built from
$(ij)\in E$: see Fig.~\ref{fig_msgs} for an illustration of these
definitions.

The basic idea of our approach is to introduce a probability law
$p({\cal C};G,u) = u^{|E_{\cal C}|}/Z(G,u)$ over the set of circuits
of $G$.  Hence the normalization factor $Z(G,u) = \sum_{\cal C}
u^{|E_{\cal C}|}$ is equal to the generating function of the number of
circuits, $\sum_{L} {\cal N}_L(G) u^L$. In the limit of large graphs
and circuit lengths, the saddle-point method leads to the relation
$\displaystyle{ \frac{1}{N}\ln Z(G,u) =\max _{\ell} [ \sigma (\ell) +
\ell \, \ln u]}$.  This relation can be inverted with standard
Legendre transformations, and the entropy $\sigma$ can be expressed in
terms of the partition function $Z$. An estimate of $Z$ can then be
obtained by using the Bethe approximation of the corresponding
statistical mechanics model, or by means of Monte Carlo
simulation~\cite{MC_enumeration}.  Following the former road, and
using the well known correspondence between minimization of the Bethe
free-energy and iterations of the Belief Propagation
equations~\cite{Yedidia}, one is lead to the following algorithm
(see~\cite{long} for details):

\begin{center}
---------------------------------------------------------------------

{\sc Circuit Counting Algorithm}
\end{center}

\noindent
{\sc Input:} a graph $G=(V,E)$, $u$ a real positive number.

\vskip .3cm
\noindent
{\sc Operation:} iterate  the set of $2|E|$ recursive equations 
\begin{equation}
y^{(T+1)}_{i \to j} = \frac{u\underset{m\in \partial i \setminus j}{\sum} 
y^{(T)}_{m \to i}}{1+ \frac{1}{2} u^2 
\underset{m\ne n}{\underset{m,n \in \partial i \setminus j}{\sum}} 
y^{(T)}_{m \to i} \ y^{(T)}_{n \to i}} \ ,
\label{eq_msgs}
\end{equation}
from a randomly chosen  initial condition $y_{i\to j}^{(0)} >0$
until it converges (within some {\em a  priori} accuracy)
to a fixed point $y^*_{i\to j} (G,u)$ .

\vskip .3cm
\noindent
{\sc Output:} estimate of the entropy $\sigma(\ell)$
of circuits of length $N\,\ell$ with
\begin{eqnarray} \label{pp}
\ell &=&\frac 1N \sum _{(ij)\in E} p_{(ij)} \ , \quad
p_{(ij)} = \frac{ u \, y^*_{i\to j}\, y^*_{j\to i}}
{1+u \, y^*_{i\to j}\, y^*_{j\to i}}\ , \\
\sigma(\ell) &=& \frac 1N \sum _{i\in V} \ln \left( 
1+ \frac{1}{2} u^2\underset{m\ne n}{\underset{m,n \in \partial i}{\sum}} 
y^*_{m \to i} \ y^*_{n \to i} \right) % \nonumber \\ &-&
-
 \frac 1N 
\sum _{(ij)\in E} \ln \big( 
1+ u \, y^*_{i \to j} \,
y^*_{j \to i} \big) - \ell \; \ln u\;. \nonumber
\end{eqnarray}

\begin{center}

---------------------------------------------------------------------
\end{center}

The procedure has to be repeated with different values of $u$ to
reconstruct a parametric plot of $\sigma(\ell)$ for $\ell \in
[0;\ell_{\rm max}]$. For small values of $u$, the iteration
equations converge to the trivial solution, $y^*=0$ for all edges. The
minimal value of $u$ yielding a non trivial solution, $u_0$, is
related to the slope of the entropy at the origin, $d\sigma/d\ell|_0=
-\ln u_0$. 

The algorithm runs in time growing polynomially with the graph size and 
logarithmically with the required accuracy on the fixed-point solution.
For generic graphs, one cannot warrant neither the convergence of the
iteration, nor the validity of the Bethe approximation (see~\cite{MoRi} for
the computation of corrections). This approximation
is however expected to be correct for large random graphs, and should be
reasonable for most real-world networks.

Besides the global information $\sigma(\ell)$, the algorithm gives a local
description of the circuits of the graph, through the quantities $y^*$, 
called {\em messages} hereafter. These have indeed the following
interpretation: for $(ij)\in E$, $p_{(ij)}$ defined in
Eq.~(\ref{pp}) is the probability that the edge is present in a
circuit ${\cal C}$ drawn from the distribution $p({\cal C}; G,u)$, {\em i.e.}
the fraction of the circuits of length $\ell$ which go through $(ij)$.

Note that in Eq.~(\ref{eq_msgs}) we used the convention that sums on
empty sets are null. In particular, $y_{i \to j} =0$ if $i$ is the
only neighbor of $j$, in other words if $i$ is a leaf of the
graph. Moreover if all incoming messages on a edge are vanishing, the
outgoing message is also null. This simple remark implies that
edges $(ij)$ with at least one of their fixed-point messages $y^*_{i
\to j}$, $y^*_{j \to i}$ vanishing are exactly the ones which would be
erased in the leaf removal procedure to compute the 2-core (maximal
subgraph in which all vertices have degree at least 2) of the graph
\cite{core}.  This property could be expected: by definition, no
circuit can be drawn outside of the 2-core.

\begin{figure}
\begin{center}
\includegraphics[width=8.3cm]{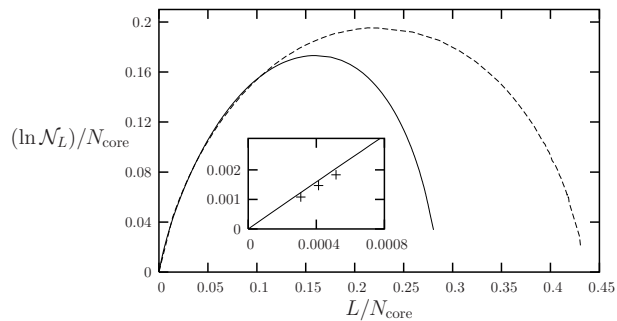}
\end{center}
\caption{Solid line: the number of circuits in a graph of the Internet.
Inset: magnification of the small length results, the symbols have been
obtained by exhaustive enumeration of the circuits of length $L=3,4,5$.
Dashed line: typical entropy in the random graph ensemble of same 
connectivity distribution.}
\label{fig_Internet}
\end{figure}

{\em Application to real-world graphs: approximate counting.}  
As an illustrative example, we present in Fig.~\ref{fig_Internet} the
output of the algorithm when applied to the graph of the
Internet structure at the Autonomous Level System, using preliminary
data from the DIMES measurement project~\cite{DIMES}. The original
graph contained $N=14291$ vertices and $M=33666$ edges. For simplicity
we plot the results in units of its $2$-core size, $N_{\rm
core}=9694$ (the $2$-core contains $M_{\rm core}=29069$ edges). Two
features of this entropy curve can be underlined.  According to our
algorithm, the most numerous circuits contain $1555$ edges, and there
are around $10^{729}$ (certainly out of reach of any direct
enumeration) of such circuits; the longest ones contain 
$L_{\rm max} \approx 2710$ edges. 
The agreement between exact enumerations for short circuits of length
$L=3,4,5$ (enumerating longer ones becomes excessively costly)
and the results of the algorithm is quantitatively 
decent (see inset of Fig.~\ref{fig_Internet}). A rough analysis of the
local information provided by the algorithm shows that high degree vertices
belongs generally to a higher fraction of circuits than poorly connected ones;
there are however strong fluctuations around this general trend.

{\em Application to random graphs: analytical results.}
Consider now random graph ensembles with fixed degree distribution, 
and call $q_k$ the fraction of vertices having degree $k$. 
We assume that $q_k$ decays fast enough for large connectivities, so
that all its moments are well defined. Let us introduce the average degree 
$c$, and the probability that the end-vertex of a randomly chosen edge has 
degree $k+1$, $\tilde q_k = (k+1) q_{k+1}/c$. The typical (quenched) entropy
density is defined by 
$\sigma_{\rm q}(\ell)= \overline{\ln {\cal N}_{\ell N}(G)}/N$, where the
over-line denotes an average over the random graph ensemble. In contrast
the computation of~\cite{BiMa} yields the annealed entropy
$\sigma_{\rm a}(\ell)= \ln \overline{{\cal N}_{\ell N}(G)}/N$.

\begin{figure}
\begin{center}
\includegraphics[width=8cm]{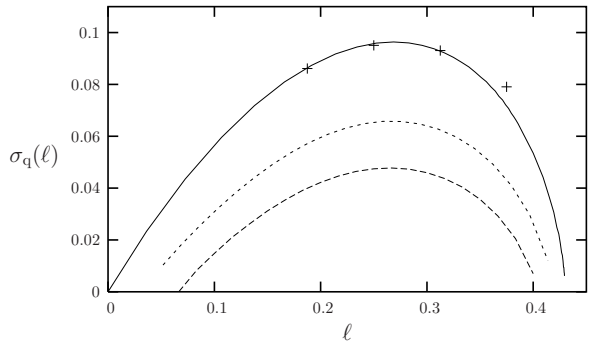}
\end{center}
\caption{Typical entropy for Poissonian graphs with $c=2$,
cavity computation (solid line) and exhaustive enumeration determination of 
the median number of cycles in samples of 200.000 graphs of size 
$N=60$ (dashed) and $N=116$ (dotted). Symbols obtained by an extrapolation
of the numerical results for several values of $N$ to the limit $N\to \infty$.}
\label{fig_poisson}
\end{figure}

Running the algorithm defined above on a graph of the ensemble leads
to a random (with respect to the choice of the graph) set of messages
$y^*$. The assumptions of the so-called cavity method at the 
replica-symmetric level~\cite{Beyond} lead to a self-consistent
equation for the distribution $P$ of the messages $y$ found on a
randomly chosen directed edge~:
\begin{eqnarray}
P(y;u) &=& \tilde q_0 \; \delta (y) +
\sum_{k=1}^\infty \tilde{q}_k \int_0^\infty \prod _{m=1}^k dy_m P(y_m;u)
\nonumber \\ &\times&
 \delta\left(y-g\big(\sum _{m=1}^k y_m,
\sum _{m=1}^k y_m^2 ,\; u\big)\right) \;,
\label{eq_Phaty}
\end{eqnarray}
where $g(a,b,u)=u a /(1+ u^2(a^2-b)/2)$ (see Eq.~(\ref{eq_msgs})).
From the solution of this distributional equation, easily found numerically
by means of a population dynamics algorithm~\cite{cavity}, one can use 
Eq.~(\ref{pp}) to compute the typical entropy of the graphs of the ensemble,
$\sigma_{\rm q}(\ell)$, parametrized by $u$. We present in 
Fig.~\ref{fig_poisson} the results of this approach on Poissonian graphs
(i.e. $q_k=e^{-c} c^k /k!$) with mean degree $c=2$, along with a confirmation 
by exhaustive enumeration on finite size samples.

An analytic resolution of Eq.~(\ref{eq_Phaty})
is possible only for the very particular case of random
regular graph for which all $q_k$ but one vanish. We find that
$P$ reduces to a single Dirac distribution in $y^*(c)$ solution
of $y^*=g\big( (c-1) y^*, (c-1) (y^*)^2,u\big)$.
In that case
the fluctuations of ${\cal N}_L$ are sufficiently small for the annealed
and quenched averages to coincide and  the result obtained rigorously
in~\cite{Garmo} is found back~\cite{MaMo}. 

Some analytical predictions can be made when the random graphs are not purely 
regular, even if $P$ is not known explicitly. 
First of all, the fraction
$\zeta$ of strictly vanishing messages is found to be the smallest root in
$[0;1]$ of  $\zeta = \sum_{k\ge 0}  \tilde {q}_k\; \zeta ^k$.
Following the above interpretation of the null messages, the
fraction of edges that belong to the 2-core is $(1-\zeta)^2$. Moreover
its connectivity distribution can also be expressed from $\zeta$ and
$\tilde{q}_k$, and
these predictions checked from the solution of the differential equations
describing the leaf removal algorithm \cite{core,long}.

One can also set up a systematic expansion of $\sigma_{\rm q}$ around 
$\ell=0$. To state the results in a compact way, let us define the
factorial moments of $\tilde{q}_k$ as
$\displaystyle{
\tilde{\mu}_n = \sum_{k\ge n}\tilde{q}_k \, k (k-1) \dots (k-n+1)}$.
The coefficients of the second order expansion of the entropy read:
\begin{equation}
\left. \frac{d\sigma_{\rm q}}{d\ell} \right| _0  = \ln \tilde{\mu}_1 
\ , \
\left. \frac{d^2\sigma_{\rm q}}{d\ell ^2} \right| _0= - \frac{1}{c} \left( 
\frac{\tilde{\mu}_3 }{\tilde{\mu}_1^2}
+ \frac{2 \tilde{\mu}_2^2 }{\tilde{\mu}_1^3 (\tilde{\mu}_1 -1)} 
\right) \ .
\end{equation}
Comparing this expansion with the annealed computation of~\cite{BiMa}, 
one finds that the first derivatives are equal in both
computations, and match the known results for circuits of finite size.
However the second derivatives turn out to be different,
\begin{equation} \label{ent2}
\left. \frac{d^2\sigma_{\rm a}}{d\ell ^2} \right| _0 -
\left. \frac{d^2\sigma_{\rm q}}{d\ell ^2} \right| _0 =
\frac{2}{c \tilde{\mu}_1^3 (\tilde{\mu}_1-1)} (\tilde{\mu}_2 - 
\tilde{\mu}_1 (\tilde{\mu}_1 -1))^2 \ .
\end{equation}
It is straightforward to show from Eq.~(\ref{ent2})
that the expansion of the annealed and quenched entropies coincide only if 
the distribution $\tilde{q}_k$ is supported by a single 
integer, in other words in the random regular graph case. 

Another limit that can be investigated analytically is the one of
maximal length circuits $\ell_{\rm max}$, reached here when $u$
gets large. We need to distinguish two cases: if the connectivity
distribution is supported on the integers larger than $3$, the cavity
computation predicts $\ell_{\rm max}=1$, and the graphs in
such ensembles are typically Hamiltonian. Interestingly, this
was conjectured by Wormald in~\cite{Wormald_review}. Even if the
present statistical mechanics approach does not provide a rigorous
proof of the conjecture, it allows to make it quantitative (with the
prediction of the typical entropy of such Hamiltonian circuits,
$\sigma_{\rm q}(1)$). Moreover it gives a hint at why usual
probabilistic methods are not powerful enough to prove the conjecture
(the quenched entropy is strictly smaller than the annealed one in
general).  Note that this property concerning Hamiltonian circuits
crucially relies on the fast decay of the degree distribution: it was
shown in~\cite{BiMa} that it can be invalidated when $q_k$ has power
law tails.

As soon as the connectivity distribution of the 2-core contains a finite 
fraction of sites of degree 2, it cannot be Hamiltonian. 
Consider indeed a vertex of degree
$k$, surrounded by $k'$ neighbors of degree 2, with $k\ge k' \ge 3$: 
it is obvious that no circuit can visit more than two of these $k'$ sites.
As the number of such forbidden vertices is extensive, one has 
$L_{\rm max}/N_{\rm core} < 1$. The quantity $\ell_{\rm max}$ can be computed
by taking analytically the appropriate limit in the equation 
(\ref{eq_Phaty}) on $P$, resulting in a simpler distributional equation 
which can be solved analytically 
in the limit of infinitesimal fraction of degree 2 sites~\cite{long}.

{\em Discussion and Conclusion.}
We mentioned in the introduction the possibility of using global properties
of graphs to test the relevance of random graph ensembles for the
description of real world networks. Following this idea, 
we compared the circuit entropy of the DIMES Internet graph with
the quenched result for the ensemble with the same connectivity distribution 
(dashed line in Fig.~\ref{fig_Internet}). They turn out to be rather 
different, suggesting that random graph ensembles defined
only through their connectivity distribution are not a very precise
description of real world networks. It would thus be interesting to 
extend our analytical study to different ensembles of graphs, for instance 
introducing correlations between the degrees of neighboring 
vertices~\cite{correlated}, or considering growing models of 
networks~\cite{growing}.

Concerning the application of the algorithm to individual graphs, two questions
should be further investigated: can one give general 
conditions~\cite{Tati,Heskes} on the 
graphs which ensure the convergence of the BP equations? Can they be 
sharpened to show that the output of the algorithm is a rigorous lower bound
on the true number of circuits? Tests on
various types of graphs show that the iteration procedure is generally
very robust against the initial condition on the $y$, and converges to a 
unique fixed-point. Small counter-examples on which the BP equations do not
converge can however be easily tailored.

Large deviations of ${\cal N}_L$ around its typical value 
$e^{N\sigma _q (\ell)}$ could also be an interesting object of study, using
for instance the modified cavity method of~\cite{rate}. One may 
in particular seek the exponentially small probability 
that a randomly drawn graph is not Hamiltonian for ensembles 
whose typical instances are.

Finally, we hope that our algorithm will be useful for
analyzing graphs data available in various contexts besides the
Internet e.g. regulatory and more generally biological interaction
networks. The huge size of these data sets make the use of exact
analysis procedures impossible, not-to-say unnecessary when data are
plagued by false positives and/or negatives as is often the case in
biological experiments e.g. DNA chips. Approximate and fast algorithms
may then reveal adequate.

{\em Acknowledgments.}  We gratefully acknowledge useful discussions
with A.~Montanari, F.~Ricci-Tersenghi and M.~Weigt. We also thank
S.~Kirkpatrick, Y.~Shavitt and E.~Shir for providing the preliminary
DIMES data.  The work was supported by EVERGROW, integrated project
No. 1935 in the complex systems initiative of the Future and Emerging
Technologies directorate of the IST Priority, EU Sixth Framework.

\end{document}